\documentclass[3p,twocolumn]{elsarticle}

\usepackage{lineno,hyperref}

\journal{Nucl.~Instr.~Meth.}

\begin{document}

\begin{frontmatter}


\title{Research and Development of Commercially Manufactured Large GEM Foils}




\author[Temple]{M. Posik\corref{mycorrespondingauthor}}
\cortext[mycorrespondingauthor]{Corresponding authors}
\ead{posik@temple.edu}

\author[Temple]{B. Surrow\corref{mycorrespondingauthor}}
\ead{surrow@temple.edu}

\address[Temple]{Temple University, Philadelphia, Pennsylvania 19122, USA}

\begin{abstract}
With future experiments proposing detectors that utilize very large-area GEM foils,
there is a need for commercially available GEM foils. Double-mask etching techniques
pose a clear limitation in the maximum size of GEM foils. In contrast, single-mask techniques developed
at CERN would allow one to overcome those limitations. However with interest in GEM foils increasing and CERN being the only main
distributor, keeping up with the demand for GEM foils will be difficult. Thus the commercialization of GEMs has been established by Tech-Etch of Plymouth, MA, USA using
single-mask techniques.

We report on the electrical and geometrical properties, along with the inner and outer hole diameter size uniformity of 10 $\times$ 10 cm$^2$ and
40$\times$40 cm$^2$ GEM foils. The Tech-Etch foils were found to have excellent electrical properties. The measured
mean optical properties were found to reflect the desired parameters and are consistent with those measured
in double-mask GEM foils, and show good hole diameter uniformity over the active area.
These foils are well suited for future applications in nuclear and particle physics where 
tracking devices are needed.
\end{abstract}

\begin{keyword}
\texttt{GEM}\sep \texttt{EIC}\sep \texttt{single-mask}\sep \texttt{tracking}\sep \texttt{micro-pattern gas detectors}\sep
\end{keyword}

\end{frontmatter}


\section{Introduction}


With the invention of the gas electron multiplier (GEM) in 1997~\cite{Sauli:1997qp}, GEM technology has attracted 
a lot interest from the nuclear and particle physics community, as they are excellent candidates to be used in tracking detectors.~GEMs have several attractive features, they function well in high rate environments (better than $10^5$ Hz/mm$^2$~\cite{Altunbas:2002ds}), have superb spatial resolution (40 $\mu m$ rms~\cite{Altunbas:2002ds}), and can cover a large acceptance. Many experiments such as STAR~\cite{Surrow:2010zza}, COMPASS~\cite{Altunbas:2002ds}, and others have already successfully incorporated GEM technology into their tracking detectors. 


 As GEM technology matures, more and more future experiments and their upgrades, including ALICE~\cite{Gasik:2014sga}, Jefferson Lab's Super Bigbite Spectrometer~\cite{SBS}, CMS~\cite{Abbaneo:2014}, and a potential EIC~\cite{EIC} are looking to build tracking detectors which take advantage of GEM technology. Future GEM detectors 
will require large area GEM foils to obtain an adequate acceptance. However, currently there are two main constraints preventing 
the use of large area GEM foils. The first is the limitation associated with the raw material. Standard raw material rolls, 
consisting of a polyimide layer sandwiched by two copper layers, have a width of about 50 cm, restricting the maximum width of the
 GEM foil. Recently this issue has been addressed through the investigation of a splicing technique~\cite{Alfansi:2010} that 
is used to splice two GEM foils together, resulting in a wider GEM foil. The second limitation facing large area GEM foils is due
to the fabrication process by which micro patterns are etched into the copper clad polyimide material, which involves the precise 
alignment of two masks (described in section~\ref{sec:double_mask}). Obtaining an accurate alignment of two masks becomes increasingly 
difficult as the GEM foil area increases, which results in limiting the area of the GEM foil. The fabrication limitation of this
{\it{double-mask}} etching technique has now been overcome by employing a {\it{single-mask}} etching 
technique, which was developed at CERN (described in section~\ref{sec:single_mask})~\cite{Villa:2010w}.

As the experimental interest and incorporation of GEM technology into future experiments grows, the need for not only readily available
large area GEM foils, but also a means by which to readout the signals produced by the GEM foils becomes essential. Currently CERN is 
the only main distributor of large area GEM foils, and will be hard pressed to keep up with the increasing demand. To help satisfy the 
GEM foil demand, the commercialization of large area GEM foils through the single-mask and double-mask processes has been established by Tech-Etch of Plymouth, MA, USA~\cite{Tech-Etch}~\cite{Surrow:2010zza,Becker:2006,Surrow:2007,Simon:2009}.
Additionally, Tech-Etch has also established a process by which to produce 2D readout foils, which are used to readout 
signals from the GEM foils.

\begin{figure}[h!]
\center
\includegraphics[width = \columnwidth, angle=0]{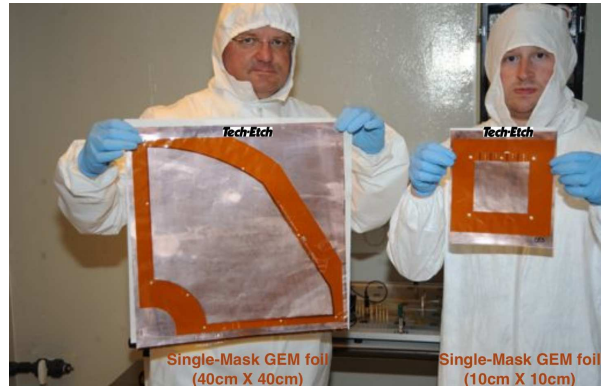}
\caption{Tech-Etch single-mask foils of areas 40$\times$40 cm$^2$ (left foil) and 10$\times$10 cm$^2$ (right foil).}
\label{fig:gem-foils}
\end{figure}

In this paper we will focus exclusively on the performance of the Tech-Etch produced $10\times10$ cm$^2$ and $\sim 40\times40$ cm$^2$ single-mask GEM foils, which can be seen in figure~\ref{fig:gem-foils}. We will begin by reviewing the production techniques in 
section~\ref{sec:gem_techniques}. Our quality control setup used to characterize and determine the integrity of each
foil is discussed in section~\ref{sec:qa}. The electrical performance and measured optical properties are presented in section~\ref{sec:performance}.
Finally, in section~\ref{sec:summary} we provide a summary and outlook concerning future developments of Tech-Etch's GEM foil production.
   
\section{Review of GEM foil production techniques}\label{sec:gem_techniques}
\subsection{Introduction}
The initial production of GEM foils employed what is known as the double-mask etching technique, which requires two masks, one on the front side and the other on the back side of the GEM foil. While this provides one with a precise micro hole pattern, as mentioned earlier this method hinders the maximum size of the GEM foil. By adopting a etching process know as the single-mask etching technique, which requires only one mask on the front side of the GEM foil, this size constraint can be alleviated.  

\subsection{Double-mask etching techniques}\label{sec:double_mask}
\begin{figure}[h!]
\center
\includegraphics[width = \columnwidth, angle=0]{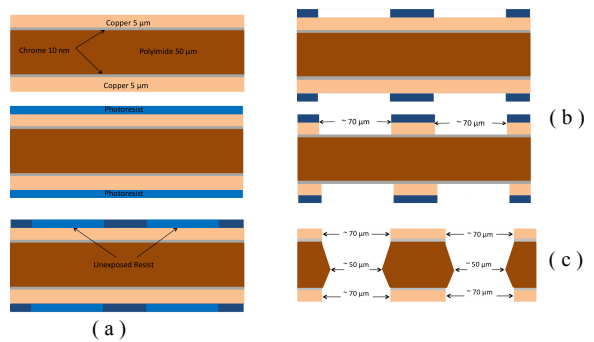}
\caption{Illustration of the double-mask etching technique. (a) shows the steps going from the bare material to the realization of the double-mask. (b) shows the hole pattern transfer to the photoresist layer via UV exposure and results of the copper etching. (c) illustrates the double conical end result of the double-mask etching technique, after the foil is submerged into the polyimide specific solvent.}
\label{fig:double-mask-etch}
\end{figure}

In the double-mask process, illustrated in figure~\ref{fig:double-mask-etch}, the bare polyimide material (about 50 $\mu m$ thick) and copper layers ($\sim$ 5 $\mu m$ each) are coated with photoresist. Laser direct imaging is then used to create the desired hole pattern on two identical masks, which are located on either side of the foil. It is important that these masks be precisely aligned, otherwise the holes on the front and back sides of the foil will be asymmetric and not uniform, which could ultimately lead to a loss in overall performance. As the area of the foil increases it becomes more difficult to precisely align both masks. The foil is then exposed to UV light, where a hole pattern is transfered through the transparent parts of the mask onto both photoresist layers. The foil is then placed in an acid bath where exposed copper is etched down to the polyimide layer. Submersion into a polyimide specific solvent with the exposed metal acting as a mask removes the exposed polyimide material from both sides, resulting in a double conical channel.

\subsection{Single-mask etching techniques}\label{sec:single_mask}
\begin{figure}[h!]
\center
\includegraphics[width = \columnwidth, angle=0]{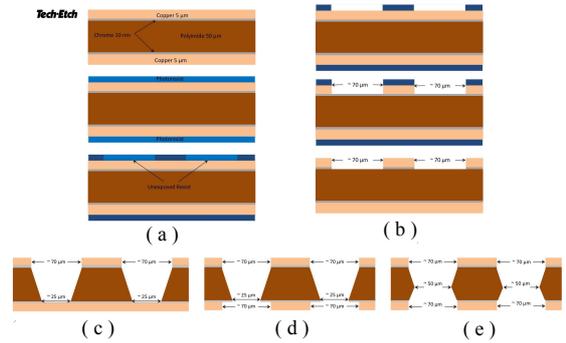}
\caption{Illustration of the single-mask etching technique. (a) shows the steps going from the bare material to the realization of the single-mask. (b) shows the hole pattern transfer to the photoresist layer via UV exposure, the removal of the unexposed photoresist, and the results of the copper etching. (c) illustrates the results after the first EDA chemistry etching is performed on the polyimide layer. (d) shows the foil after the back side copper layer is electrolytically etched, resulting in hole diameters similar to that of the front side. (e) illustrates the final double conical structure of the holes, after a second EDA chemistry etching of the polyimide layer.}
\label{fig:single-mask-etch}
\end{figure}

\begin{figure}[h!]
\center
\includegraphics[width = 0.5\columnwidth, angle=90]{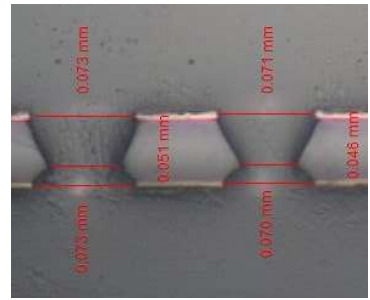}
\caption{Cross section image of a 40$\times$40 cm$^2$ single-mask Tech-Etch produced GEM foil.}
\label{fig:gem-crosssection}
\end{figure}

In order to overcome the size limitation of using the double-mask process, a single mask process can be adopted, which is highlighted in figure~\ref{fig:single-mask-etch}. The bare copper layers are first coated with photoresist. Laser direct imaging is then used to create the desired micro hole pattern on  a single mask which is placed on the front of the foil. UV light is then used to transfer the hole pattern to the photoresist layer. The unexposed photoresist is then developed away and the front layer of copper is then etched via an acid bath. The polyimide is then etched in ethylenediamine (EDA) chemistry. Following this the backside copper layer is then electrolytically etched resulting in a conical channel. To achieve a double conical channel a subsequent polyimide EDA etching is performed. For comparison, a cross section image of a Tech-Etch produced 40$\times$40 cm$^2$ single-mask GEM foil is shown in figure~\ref{fig:gem-crosssection}, in which the double conical structure of the holes can be seen.

In the past Tech-Etch had chosen to produce their double-mask GEM foils using a Kapton based polyimide material. When starting their transition to producing single-mask GEM foils it was suggested by CERN to switch the polyimide material to Apical, which resulted in better electrical performance of the GEM foils (discussed in section~\ref{sec:electrical-performance}). Both CERN and Tech-Etch now exclusively use Apical as their polyimide material. 

Tech-Etch, working in collaboration with CERN has now firmly established a commercial GEM foil fabrication process for foils up to $40\times40$ cm$^2$ using the single-mask technique.   

\section{Quality control measures}\label{sec:qa}
Quality control measures are critical for establishing a stable and reliable process by which to commercially manufacture GEM foils. There are two main areas of characterization that can be performed on the GEM foils. The first is through characterizing the foil's electrical performance. The second area of characterization is through the foils geometric properties obtained via an optical analysis.

These analyses, carried out by Temple University, were critical towards establishing the commercial fabrication of single-mask GEM foils, as they provided direct feedback to Tech-Etch who could then modify their manufacturing process as needed in order to quickly converge on a reliable production process.

\subsection{Electrical tests}\label{sec:qa_electrical}
\begin{figure}[h!]
\center
\includegraphics[width = 0.65\columnwidth, angle=0]{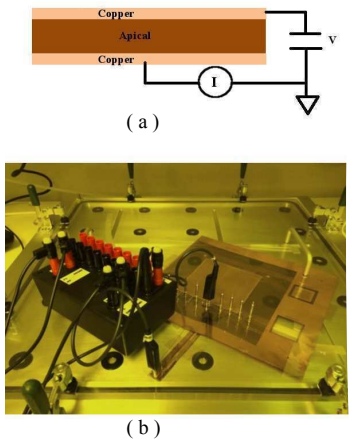}
\caption{Leakage current measurement setup. (a): Schematic of the setup to measure the leakage current of the GEM foils. (b): Image of a $10\times10$ cm$^2$ GEM foil enclosed in a nitrogen box ready for leakage current measurements.}
\label{fig:leakcurrent-setup}
\end{figure}

The electrical integrity, or the electrical insulation, of a GEM foil is determined by measuring its leakage current, shown schematically in figure~\ref{fig:leakcurrent-setup} (a). A voltage is applied across the foil and the resulting current is measured. For an ideal GEM foil, which has a resistance of 500 G$\Omega$ (calculated assuming an area of 100 $cm^2$, a typical Apical volume resistivity of $\ge$ 10$^{16}$ $\Omega\cdot$ cm~\cite{Apical_Resistivity}, and thickness of 50 $\mu m$) and a voltage of 500 V applied across the foil, one would expect a leakage current of about $\le$ 1 nA. The leakage current measurements are done in a class 1,000 clean room. The GEM foil is placed inside a plexiglass enclosure, as shown in figure~\ref{fig:leakcurrent-setup} (b), which is continuously flushed with nitrogen gas in order to provide a safe dry environment and minimize sparking. A voltage is applied between the front and back sides of the foil and slowly ramped up from 0 to 600 V, where the leakage current is then measured in 100 V increments. A ISEG SHQ 222M high voltage power supply is used to apply the high voltage and measure the resulting current. While the 10$\times$10 cm$^2$ foils have only one high voltage segment, the 40$\times$40 cm$^2$ foils are divided into nine high voltage segments.   

\subsection{Optical analysis}\label{sec:qa_optical}
\begin{figure}
\center
\includegraphics[width = \columnwidth, angle=0]{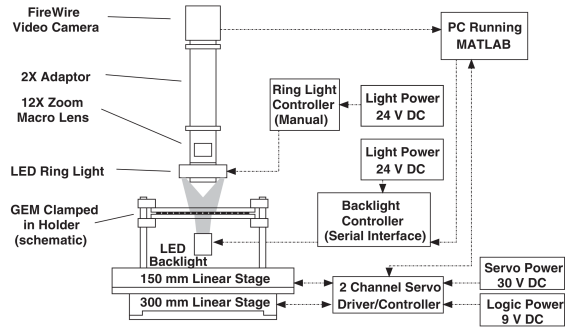}
\caption{Optical scanning setup used to analyze GEM foils. Image reproduced from~\cite{Becker:2006}.}
\label{fig:optical-scan-setup}
\end{figure}

\begin{figure}
\center
\includegraphics[width = \columnwidth, angle=0]{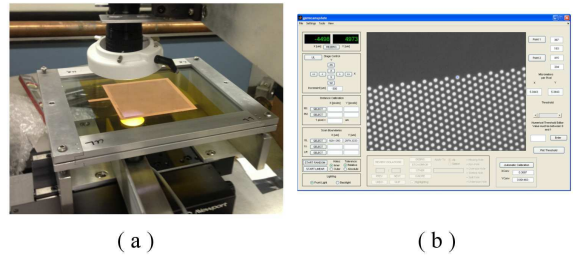}
\caption{(a) Optical scanner setup for 10$\times$10 cm$^2$ GEM foils. (b): Image of inner hole diameters on MATLAB software.}
\label{fig:ccd-setup}
\end{figure}

The optical analysis of the GEM foils is performed using an automated two dimensional CCD camera scanner. This setup is identical to that described in ref.~\cite{Becker:2006} (shown here in figure~\ref{fig:optical-scan-setup}). The CCD camera setup, shown in figure~\ref{fig:ccd-setup} (a), consists of a video camera, 2x adapter, and 12x zoom lens with a ring of LEDs around the lens face (front light). This CCD camera setup is coupled to a 2D support stage which allows for the  entire GEM foil to be scanned with high precision. The GEM foil is enclosed between two glass plates, which are secured by an aluminum frame, and is placed over an LED light (back light). The apparatus is controlled through a MATLAB~\cite{Matlab} graphical interface, shown in figure~\ref{fig:ccd-setup} (b).

\begin{figure}
\center
\includegraphics[width = \columnwidth, angle=0]{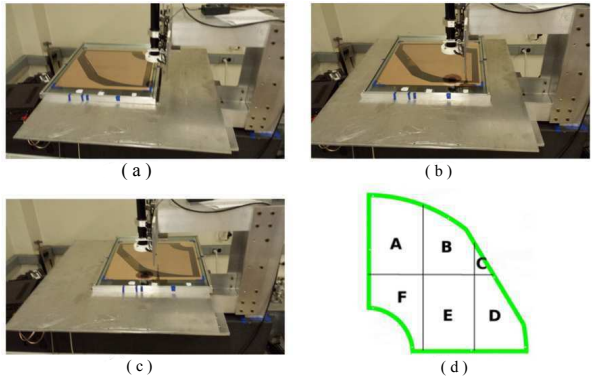}
\caption{Optical analysis setup to scan larger area GEM foils. (a)-(c): Show the three locations relative to the CCD camera that the GEM foil needs to be positioned in order to cover scan regions A-C, respectively. To cover scan regions D-F, the foil would need to be rotated by 180$^\circ$ and repositioned at each of the three specific locations shown. (d): Division of 40 $\times$ 40 cm$^2$ foil into six scan regions.}
\label{fig:fgt-ccd-setup}
\end{figure}

The stage area in the optical setup shown in figure~\ref{fig:ccd-setup} can be increased to accommodate the scanning of larger GEM foils (up to about 50$\times$50 cm$^2$) by installing a 86 $\times$ 76 cm$^2$ aluminum plate over the 2D motorized support.  
The GEM foils are again placed between two glass plates secured by an aluminum frame. Although the stage area is increased in the modification, the distance that each stage can travel ($\sim$ 15 cm in one direction and $\sim$ 30 cm in the other direction) can not cover the entire foil in one scan. Therefore the foil needs to be divided into six different scan regions labeled A through F, shown in figure~\ref{fig:fgt-ccd-setup} (d). In order to cover each scan region the foil needs to be manually moved to one of three locations relative to the CCD camera. Figure~\ref{fig:fgt-ccd-setup} (a)-(c) shows the three specific locations that relate to scan regions A-C, respectively. To scan regions D-F, the foil needs to be rotated by 180$^\circ$ and then again positioned at each of the locations shown in figure~\ref{fig:fgt-ccd-setup}.   

\begin{figure}
\center
\includegraphics[width = 0.75\columnwidth, angle=0]{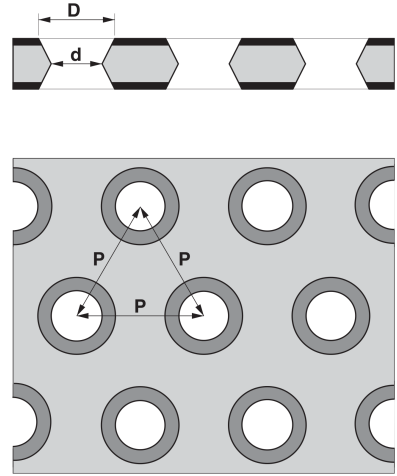}
\caption{Schematic view of Tech-Etch single-mask GEM foil. Image reproduced from ref.~\cite{Becker:2006}.}
\label{fig:foil-geometery}
\end{figure}

Through the optical analysis we are able to measure the geometrical properties of the GEM foils. Figure~\ref{fig:foil-geometery} shows three geometrical quantities that can be optically measured: the pitch (P), inner (d), and outer hole (D) diameters. The pitch is a measure of the distance between the hole centers of two neighboring holes. Performing the optical scans with the front light on and the back light off, we are sensitive to the outer hole diameters, which are defined by the copper layer and appear black in the binary converted CCD image. If the scan is performed with the front light off and back light on, the optical setup measures the inner hole diameters of the foil, which are defined by the Apical layer and appear white in the binary converted CCD image. Matlab is used to convert the pixel information obtained from the CCD binary images into measurements of the pitch, inner, and outer hole diameters.

\section{Performance of commercially produced GEM foils}\label{sec:performance}
Tech-Etch has sent Temple University three manufacturing lots of 10 $\times$ 10 cm$^2$ single-mask GEM foils for testing and characterization  with each lot containing 6/12/6 foils respectively. The first two batches varied in quality and consistency as Tech-Etch was still tuning their fabrication process. The last batch of 6 GEM foils were found to meet the specified parameters finalizing the commercial fabrication process for the 10 $\times$ 10 cm$^2$ single-mask GEM foils. The next step was then to establish a commercial fabrication procedure for larger (40 $\times$ 40 cm$^2$) single-mask GEM foils. To verify the quality and consistency of these larger foils, Tech-Etch has sent Temple University three foils for testing. Measurements of these larger foils were found to be consistent with those of the 10 $\times$ 10 cm$^2$ foils, thus establishing the commercialization of large area single-mask GEM foils.   

\subsection{Electrical performance}\label{sec:electrical-performance}
\begin{figure}
\center
\includegraphics[width = \columnwidth, angle=0]{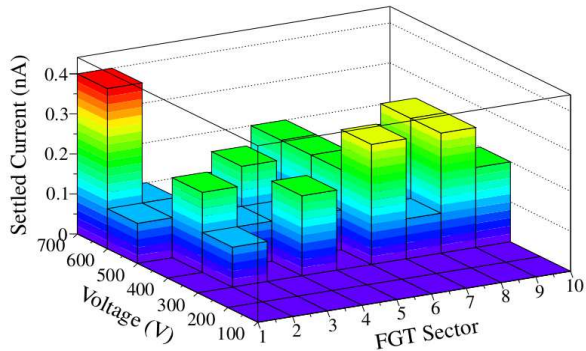}
\caption{Measured leakage current as a function of voltage and sector (1-9) for a single-mask 40 $\times$ 40 cm$^2$ GEM foil. The measured current is accurate to within 0.5 nA.}
\label{fig:fgt-leakage-current}
\end{figure}

All single-mask Tech-Etch GEM foils (10 $\times$ 10 cm$^2$ and 40 $\times$ 40 cm$^2$) displayed excellent electrical performance. The typical leakage current, measured using the setup described in section~\ref{sec:qa_electrical}, was consistently less than about 1 nA. These results were independently verified by Tech-Etch prior to shipping the foils to Temple University. This is much improved from the previous Tech-Etch produced double-mask GEM foils, which typically saw a leakage current on average around 10 times larger. The improvement in the electrical performance is due to switching the insulating material from Kapton to Apical, which has a lower water absorption rate~\cite{Kapton,Apical}. Additionally, three single-mask GEM foils produced at CERN (with the polyimide layer made from Apical) were also tested and consistently displayed leakage currents below 1 nA. Figure~\ref{fig:fgt-leakage-current} shows the leakage current for a 40 $\times$ 40 cm$^2$ single-mask GEM foil, produced by Tech-Etch, as a function of applied voltage across the foil for each of the foils nine sectors. These results were typical of all measured single-mask GEM foils.

\subsection{Optical performance}\label{sec:optical-performance}
The geometrical properties of the single-mask GEM foils are measured using the optical analysis setup described in section~\ref{sec:qa_optical}. These properties include the pitch between each hole, the inner and outer hole diameters, and their uniformity across the entire active area of the GEM foil. This type of optical analysis is critical towards establishing a consistent commercial fabrication process, as inconsistencies between these parameters, or their non-uniformity across the GEM foil will degrade the performance of the GEM. An exhaustive optical analysis of the two different single-mask GEM foil sizes (10 $\times$ 10 and 40 $\times$ 40 cm$^2$) was carried out and is presented in the following subsections.

\subsubsection{10 $\times$ 10 cm$^2$}\label{sec:optical-performance-10x10}
\begin{figure}
\center
\includegraphics[width = \columnwidth, angle=0]{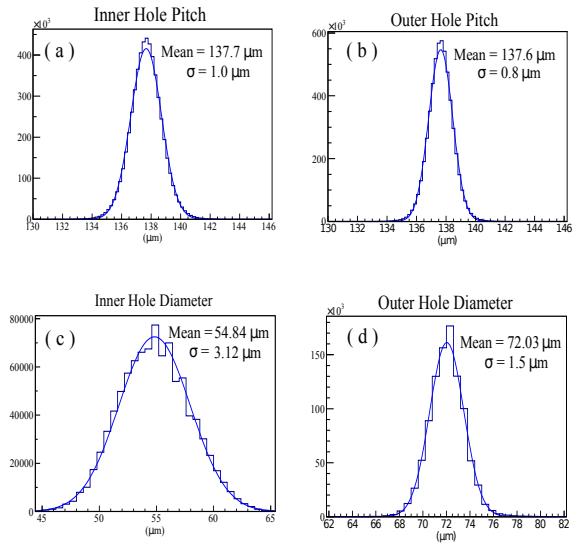}
\caption{Representative sample of the single-mask 10 $\times$ 10 cm$^2$ GEM foils geometrical distributions. (a): Pitch distribution measured using the back light. (b): Pitch distribution measured using the front light. (c): Inner hole diameter distribution. (d) Outer hole diameter distribution.}
\label{fig:10x10-distribution}
\end{figure}

The last lot of six 10 $\times$ 10 cm$^2$ single-mask GEM foils showed similar properties amongst themselves. Figure~\ref{fig:10x10-distribution} shows a representative sample of one of the 10 $\times$ 10 cm$^2$ single-mask GEM foils produced by Tech-Etch. Figure~\ref{fig:10x10-distribution} (a) displays a typical pitch distribution measured while back lighting the GEM foil (sensitive to the inner hole diameters), while figure~\ref{fig:10x10-distribution} (b) shows the pitch distribution measured while lighting the GEM foil from the front (sensitive to the outer hole diameters). The Pitch measured using the inner or outer diameters was found to be consistent at about 138 $\mu m$. The inner hole diameter distribution is plotted in figure~\ref{fig:10x10-distribution} (c), while the outer hole distribution can be seen in Figure~\ref{fig:10x10-distribution} (d). Comparing these distributions on average over all six 10 $\times$ 10 cm$^2$ single-mask foils, the pitch had the narrowest distribution ($\sigma \approx$ 1 $\mu m$), and the inner hole diameter distribution was consistently wider ($\sigma \approx$ 2.5 $\mu m$) than the outer hole diameter distribution ($\sigma \approx$ 1.5 $\mu m$). The difference in the spread between the inner and outer hole diameter distributions is believed to be due to the Apical layer being more sensitive to the etching time than the copper layer, thus one can better control the precision of the outer diameter better. This is a general behavior of all of the single-mask GEM foils from the last manufacturing lot that we received from Tech-Etch.    

\begin{figure}
\center
\includegraphics[width = \columnwidth, angle=0]{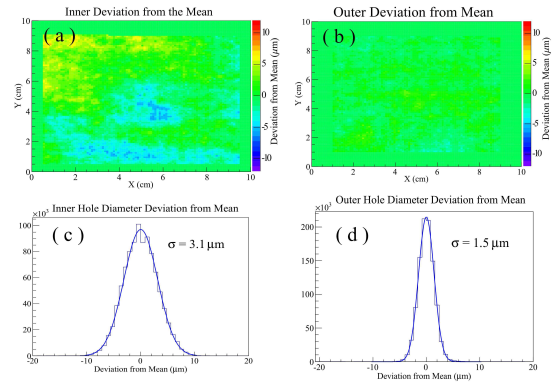}
\caption{Representative sample of single mask 10 $\times$ 10 cm$^2$ GEM foils hole diameter uniformity. (a): Inner hole diameter uniformity. (b): Outer hole diameter uniformity. (c): Inner hole diameter deviation from mean. (d): Outer hole diameter deviation from mean.}
\label{fig:10x10-uniformity}
\end{figure}

Not only is it important to know the geometrical values of the GEM foil, but also how uniform the inner and outer hole diameters are throughout the GEM foil. Figure~\ref{fig:10x10-uniformity} shows the inner (a) and outer (b) hole diameter deviation from the diameter mean over the entire GEM foil active area. The inner diameters are generally not as uniform as the outer diameters. Typical deviations of the inner hole diameters cover a range of approximately $\pm$7 $\mu m$. When fitting a Gaussian function to the inner (outer) hole diameter deviation from the mean distribution, the average sigma of the deviation is only about 3$\mu m$ (1.5$\mu m$), as can be seen in figure~\ref{fig:10x10-uniformity} (c) (figure~\ref{fig:10x10-uniformity}(d)).

\begin{figure}[htbp]
\includegraphics[width=\columnwidth]{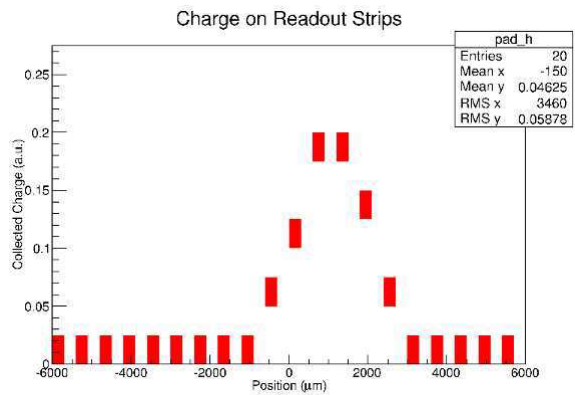}
\caption{The amount of charge collected on a simple read out strip geometry.} 
\label{fig:charge_pad}
\end{figure}

 If the hole diameter size varies widely across the foils, then this will lead to different amounts of charge being produced as the initial electron passes through or collides into (when using multiple GEM foils) GEM holes of different sizes in each of the foils. To help quantify how sensitive the track reconstruction is to the charge variation, a simple track reconstruction exercise was carried out. In this exercise it was assumed that the charge produced from a GEM foil was read out by 20 read out strips that were $520\,\mu m$ in length and each strip had a pitch of $600\,\mu m$ and was separated by 80 $\mu m$. A random Gaussian shaped charge cloud was then generated at a position $x$ within the range covered by the read out strips (about $-6000\,\mu m \le x \le 6000\,\mu m$). Figure~\ref{fig:charge_pad} shows the amount of charge collected from the charge cloud in each read out strip. The reconstructed position of the particle, is a charge weighted sum given by

\begin{equation}\label{eq:xpos}
\left< x \right> = \frac{\displaystyle \sum_{i} x_i Q_i}{ \displaystyle \sum_{i} Q_i },
\end{equation}  

\noindent where $x_i$ is the position and $Q_i$ is the charge of the $i^{\mathrm{th}}$ read out strip. The reconstructed position sensitivity to the charge was quantified by randomly varying the collected charge on each read out strip by $\pm 5$\% and $\pm 50$\%. The reconstructed position was found not to rely too much on the charge fluctuations, as the change in $\left< x \right>$ between the large charge variations of $\pm 5$\% and $\pm 50$\% was only a few $\mu m$. The most sensitive quantity to the charge variations was the resolution of the reconstructed position. Although even this was found not to be that significant overall, with the $\pm 5$\% ($\pm 50$\%) charge variation producing about a 1\% (4\%) increase in the reconstructed width relative to a reconstructed position with no charge variation.

\begin{figure}
\center
\includegraphics[width = \columnwidth, angle=0]{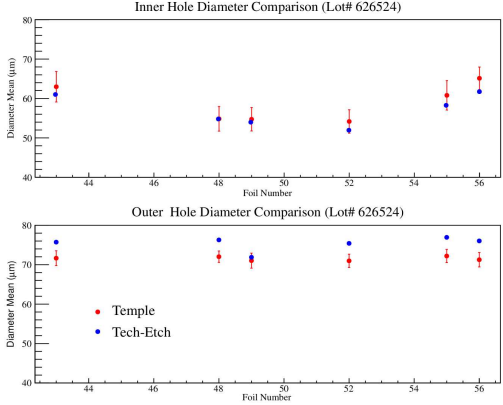}
\caption{Average inner (top panel) and outer (bottom panel) hole diameters across last batch of Tech-Etch 10 $\times$ 10 cm$^2$ single-mask GEM foils measured by Temple University (red markers) and Tech-Etch (blue markers).}
\label{fig:10x10-comp}
\end{figure}



The mean inner (outer) hole diameters for each of the six 10 $\times$ 10 cm$^2$ single-mask GEM foils are shown in figure~\ref{fig:10x10-comp}, and compared with measurements done at Tech-Etch. The error bar associated with Temple University's measurement is the RMS of the measured inner (outer) hole diameter distribution. The Tech-Etch measured diameters were performed very differently from Temple University's method. Tech-Etch performed there measurements using high resolution images of about nine holes and have no error bars shown. Nonetheless, Tech-Etch's measurements are found to be consistent with Temple University's and serve as an excellent cross check between the two measurement techniques.  

\begin{figure}[!h]
\centering
\includegraphics[width=\columnwidth]{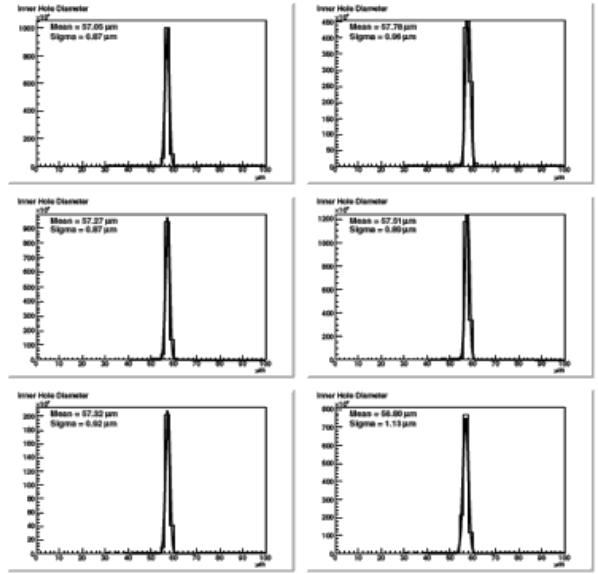}
\caption{Inner diameter distribution for a Tech-Etch 40 $\times$ 40 cm$^2$ double-mask foil split into 6 CCD scan regions. The mean inner diameter is about 57 $\mu m$ with a $\sigma$ of $\sim 1 \mu m$ for all scan regions.}
\label{fig:double-mask-ID}
\end{figure}

\begin{figure}[!h]
\centering
\includegraphics[width=\columnwidth]{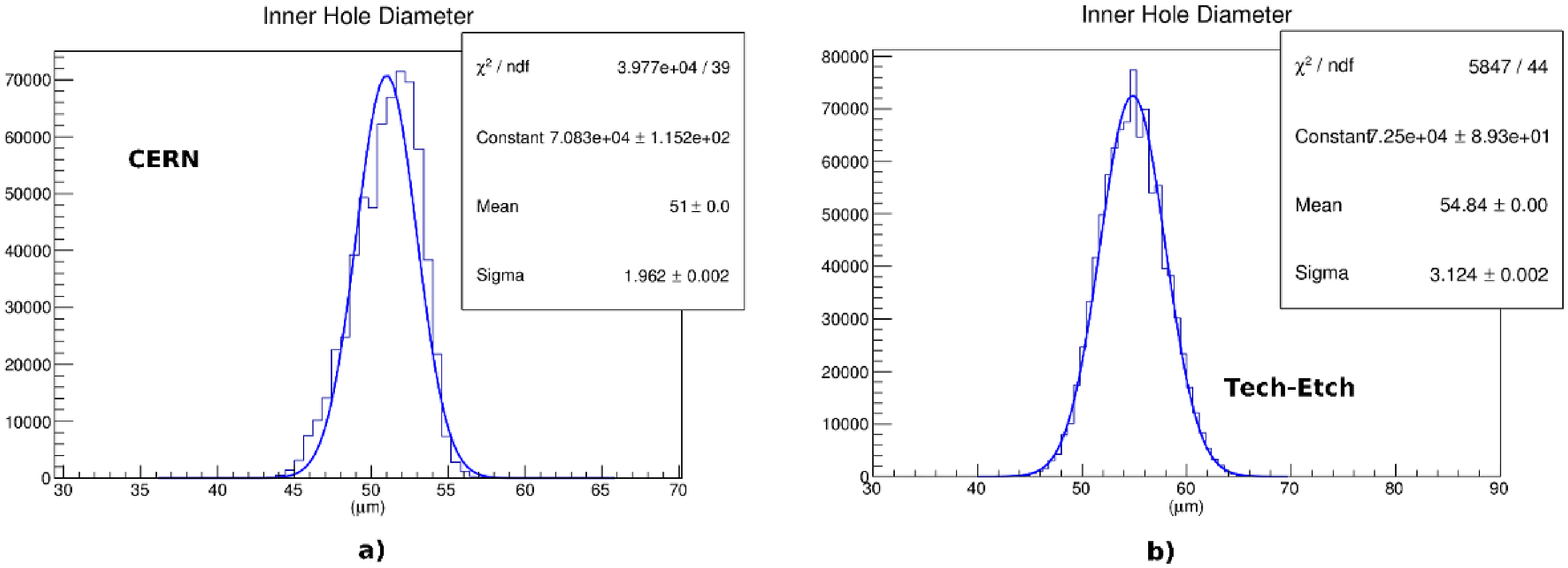}
\caption{Inner diameter distribution. a) CERN foil: Gaussian fit mean $\sim 51.0 \mu m$, $\sigma \sim 2.0 \mu m$. b) Tech-Etch foil: Gaussian fit mean $\sim 55.0 \mu m$, $\sigma \sim 3.1 \mu m$.}
\label{fig:cern-diam}
\end{figure}

\begin{figure}[!h]
\centering
\includegraphics[width=\columnwidth]{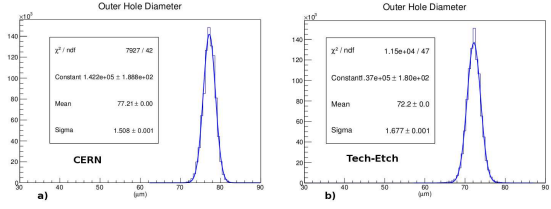}
\caption{Outer diameter distribution. a) CERN foil: Gaussian fit mean $\sim 77.2 \mu m$, $\sigma \sim 1.5 \mu m$. b) Tech-Etch foil: Gaussian fit mean $\sim 72.2 \mu m$, $\sigma \sim 1.7 \mu m$.}
\label{fig:cern-outdiam}
\end{figure}

\begin{figure}[!h]
\centering
\includegraphics[width=\columnwidth]{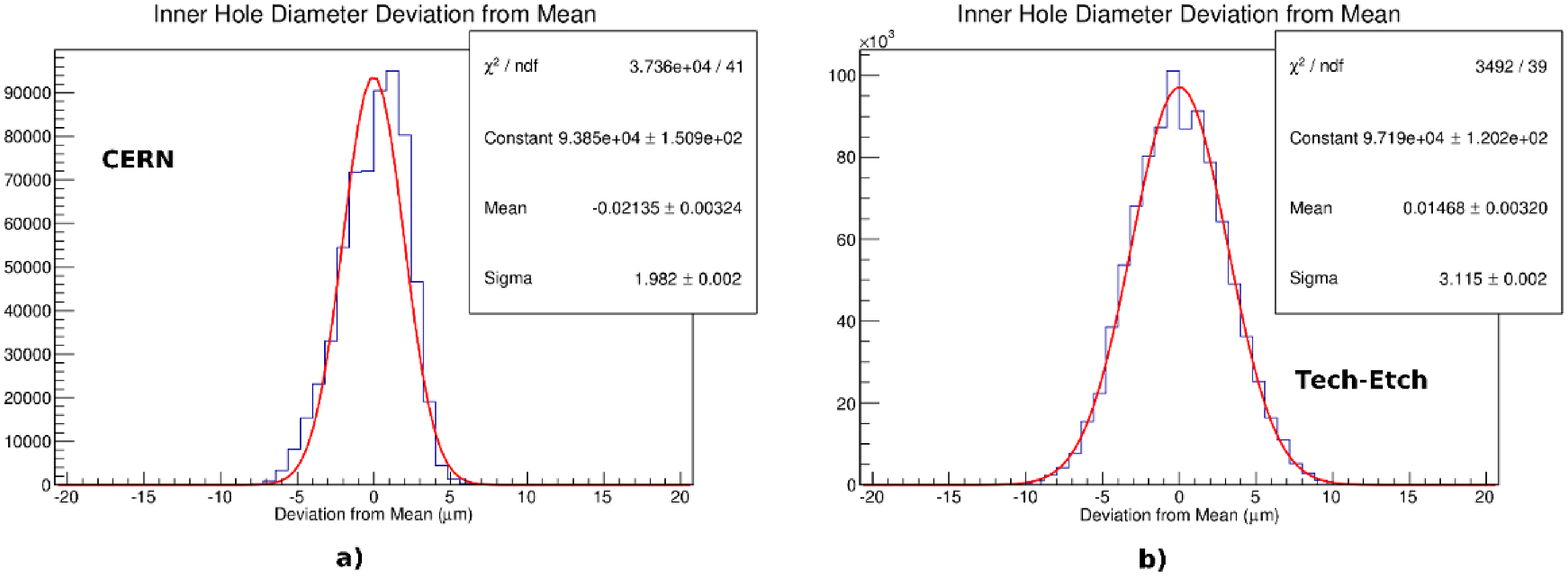}
\caption{Inner diameter deviation from the mean distribution. a) CERN foil: Gaussian fit $\sigma \sim 2.0 \mu m$. b) Tech-Etch foil: Gaussian fit $\sigma \sim 3.1 \mu m$.}
\label{fig:cern-diam-dev}
\end{figure}

\begin{figure}[!h]
\centering
\includegraphics[width=\columnwidth]{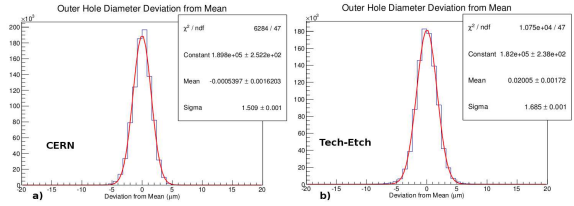}
\caption{Outer diameter deviation from the mean distribution. a) CERN foil: Gaussian fit $\sigma \sim 1.5 \mu m$. b) Tech-Etch foil: Gaussian fit $\sigma \sim 1.7 \mu m$.}
\label{fig:cern-out-diam-dev}
\end{figure}

From foil to foil, the inner and outer diameters are consistent with each other. The means from each foil are then fit with a constant line to obtain an overall average inner (outer) hole diameter across all six foils of about 58 $\mu m$ (71 $\mu m$). The pitch of the holes was found to be relatively flat across all six foils, with an overall average of about 138 $\mu m$. These geometrical parameters are consistent with those previously used in double-mask GEM foils~\cite{Becker:2006,Simon:2007sk}. Figure~\ref{fig:double-mask-ID} shows the inner diameter distributions of a double-mask 40 $\times$ 40 cm$^2$ Tech-Etch foil that have been measured at Temple University. The mean inner diameter of each distribution is about 57 $\mu m$, in good agreement with the Tech-Etch single mask diameter size. Note that the narrower distribution in the double-mask measurements ($\sigma \sim 1 \mu m$) is a result of the etching precision in the double-mask technique. The double-mask foils produced by Tech-Etch use glass tooling, whereas the single-mask foils are produced using mylar. The pitches in the double-mask measurements were found to be around 139 $\mu m$, in good agreement with what is found in the single-mask measurements. As an additional comparison, the inner/outer hole diameters and pitches of three 10$\times$10 cm$^2$ single-mask GEM foils produced at CERN were compared to its Tech-Etch counter parts. A comparison of the inner (outer) diameter and its deviation from the mean between CERN and Tech-Etch foils has also been made. Figure~\ref{fig:cern-diam} (Figure~\ref{fig:cern-outdiam}) shows the inner (outer) hole distribution for a representative CERN and Tech-Etch foil. The mean inner (outer) diameter size between the two foils are similar to one another. The Inner (outer) hole diameter deviation from the mean is shown in fig.~\ref{fig:cern-diam-dev} (fig.~\ref{fig:cern-out-diam-dev}). From these figures it can be seen that the CERN foil shows a more uniform inner hole diameter than the Tech-Etch foil, about 1 $\mu m$ narrower. Whereas the deviations in the outer hole diameter between the two foils are similar. Finally, the pitch distributions between the CERN and Tech-Etch foils were found to be similar to on another in both mean ($\sim 138 \mu m$) and spread ($\sigma < \sim 2 \mu m$).

\subsubsection{40 $\times$ 40 cm$^2$}\label{sec:optical-performance-40x40}
The optical analysis of the larger 40$\times$40 cm$^2$ followed the same procedure used to measure the geometrical properties of the 10$\times$10 cm$^2$ foils. However, whereas all of 10$\times$10 cm$^2$ foil images were taken with one CCD scan, the 40$\times$40 cm$^2$ foils needed to be divided into six CCD scan regions due to the translational limitation of our 2D stage, as shown in figure~\ref{fig:fgt-ccd-setup} (d). 

\begin{figure}[!h]
\centering
\includegraphics[width=\columnwidth]{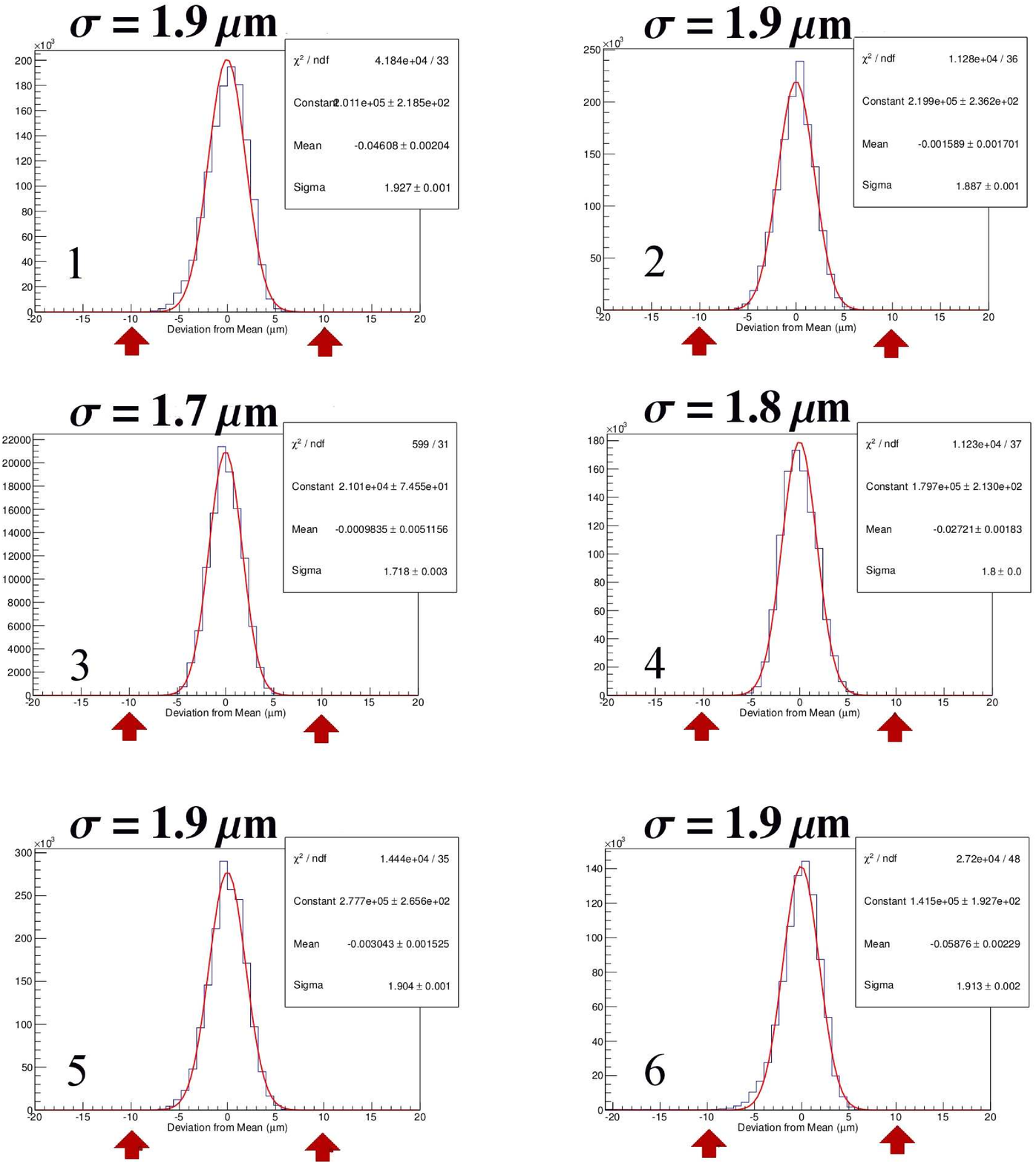}
\caption{Inner hole diameter deviation from mean for CCD scan regions 1-6. The red arrows mark the $\pm$10 $\mu m$ position.}
\label{fig:fgt-uniformity}
\end{figure}

Similar distributions (pitch, inner, and outer hole diameters) were measured with the 40$\times$40 cm$^2$ foils as were found in the 10$\times$10 cm$^2$ foils. Many of the same geometrical behaviors found in the 10$\times$10 cm$^2$ were also seen in the larger foils. In particular the pitch displayed the narrowest distribution and the inner hole diameters showed a larger deviation from the mean than the outer hole diameters. Also like the 10$\times$10 cm$^2$ foils, the hole diameters were found to have excellent uniformity across the 40$\times$40 cm$^2$ foils, where deviations were found to be smaller $\pm10 \mu m$, as shown in fig.~\ref{fig:fgt-uniformity}. The inner (outer) hole diameter deviation distribution widths generally ranged from $\sigma$ = $1.7$ to $3.0 \mu m$ ($\sigma$ = $1.1$ to  $1.8 \mu m$).      

\begin{figure}[!h]
\centering
\includegraphics[width=\columnwidth]{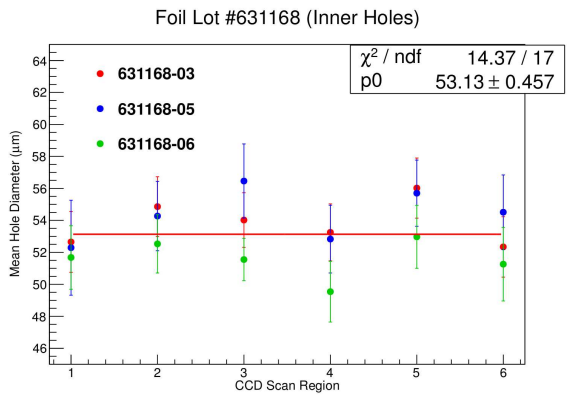}
\caption{Average inner hole diameters for each CCD scan region for all three 40$\times$40 cm$^2$ foils. The error bars represent the sigma of a Gaussian fit to that particular distribution. A constant line is fit to get the mean inner hole diameter across all of the foils.}
\label{fig:fgt-id}
\end{figure}

\begin{figure}[!h]
\centering
\includegraphics[width=\columnwidth]{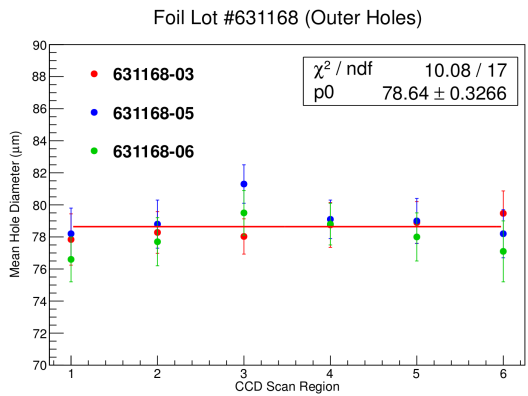}
\caption{Average outer hole diameters for each CCD scan region for all three 40$\times$40 cm$^2$ foils. The error bars represent the sigma of a Gaussian fit to that particular distribution. A constant line is fit to get the mean outer hole diameter across all of the foils.}
\label{fig:fgt-od}
\end{figure}

Considering all three of the 40$\times$40 cm$^2$ foils, we measured a near constant pitch of about 138 $\mu m$ in each CCD scan region across all foils. The average inner (outer) hole diameters were found to be consistent over all CCD scan regions across all three foils, as shown in fig.~\ref{fig:fgt-id} (fig.~\ref{fig:fgt-od}). The mean inner (outer) hole diameter across all three foils was measured to be 53.13 $\mu m$ (78.64 $\mu m$), which are similar to the double-mask GEM foil values found in ref.~\cite{Becker:2006}.   

\section{Summary and outlook}\label{sec:summary}
Tech-Etch has successfully manufacture single-mask GEM foils of 10 $\times$ 10 and 40 $\times$ 40 cm$^2$. Through electrical and optical analysis of the GEM foils, it has been determined that they meet the requirements needed by the nuclear and particle physics community to be used in tracking detectors; thus successfully establishing the commercialization of single-mask GEM foils. The electrical performance of these foils was superb, due to the changing of the polyimide layer from Kapton to Apical. The optical properties of these foils were found to be consistent with previously measured double-mask GEM foils and single-mask foils produced by CERN. 

With the recent successful production of 50 $\times$ 50 cm$^2$ single-mask GEM foils, Tech-Etch now plans on extending this process to even larger (on the order of 1 m$^2$ long foils) single-mask GEM foils, and are currently in the process of upgrading their fabrication facilities in order to handle the larger area GEM foils. 

The commercialization of the large area single-mask GEM foils will go a long way towards alleviating some of the demand for large GEM foils.

\section{Acknowledgment}
We would like to thank David Crary, Kerry Kearney, and Matthew Campbell (Tech-Etch Inc.), as well as M.~Hohlmann (FIT), R.~Majka (Yale), and especially R.~De~Oliveira (CERN) for their useful discussions, guidance, and expertise which has lead to the successful commercialization of GEM technology. This research was funded by an EIC R\&D grant, subcontracted by Brookhaven Science Associates, \#223228.


@article{Sauli:1997qp,
  journal = "Nucl.\ Instrum.\ Meth.\ A ",
  volume  = "386",
  pages  = "531",
  year    = "1997",
  author  = "F.~Sauli"
}

@article{Becker:2006,
      journal        = "Nucl.\ Instrum.\ Meth.\ A",
      volume         = "556",
      pages          = "527-534",
      year           = "2006",
      author         = "U.~Becker and B.~Tamm and S.~Hertel"
}

@article{Villa:2010w,
   author = "M.~Villa and S.D.~Pinto and M.~Alfonsi and I.~Brock 
             and G.~Croci and E.~David and R. de Oliveria and L.~Ropelewski {\it{et al.}}",
   journal = "Nucl.\ Instrum.\ Meth.\ A",
   volume = "628",
   pages = "182",
   year = "2011"
}

@manual{Tech-Etch,
   organization = "Tech-Etch Inc.",
   address = "45 Aldrin road, Plymouth, MA 02360",
   url = "<http://www.tech-etch.com/>."
}

@manual{Matlab,
   organization = "The Math Works Inc.",
   address = "Natick, MA",
   url = "<http://www.mathworks.com/>."
}

@article{Simon:2007sk,
      author         = "Simon, F. and Azmoun, B. and Becker, U. and Burns, L. and
                        Crary, D. and others",
      title          = "{Development of tracking detectors with industrially
                        produced GEM foils}",
      journal        = "IEEE Trans.Nucl.Sci.",
      volume         = "54",
      pages          = "2646-2652",
      year           = "2007"

}


@article{Alfansi:2010,
     journal         = "Nucl.\ Instrum.\ Meth.\ A",
     volume          = "617",
     pages           = "151",
     year            = "2010",
     author          = "M.~Alfansi and {\emph{et al.}}"
} 



@article{Altunbas:2002ds,
   author = "M.C.~Altunbas and {\emph{et al.}}",
   journal = "Nucl.\ Instrum.\ Meth.\ A",
   volume = "490",
   pages = "177",
   year = "2002"
}

@article{Surrow:2010zza,
  author = "B.~Surrow",
  journal = "Nucl.\ Instrum.\ Meth.\ A",
  volume = "617",
  pages = "196",
  year = "2010"
}

@article{Surrow:2007,
  author = "B.~Surrow and {\emph{et al.}}",
  journal = "Nucl.\ Instrum.\ Meth.\ A",
  volume = "572",
  pages = "201",
  year = "2007"
}

@article{Simon:2009,
  author = "F.~Simon and {\emph{et al.}}",
  journal = "Nucl.\ Instrum.\ Meth.\ A",
  volume = "598",
  pages = "432",
  year = "2009"
}

@article{Gasik:2014sga,
  author = "P.~Gasik",
  journal = "JINST",
  volume = "9",
  pages = "C04035",
  year = "2014"
}

@article{SBS,
  author = "K.~Gnanvo and N.~Liyanage and V.~Nelyubin and K.~Saenboonruang and Seth~Sacher and B.~Wojtsekhowski",
  journal = "arXiv:1409.5393",
  note = "(2014)."
}

@article{Abbaneo:2014,
  author = "D.~Abbaneo and {\emph{et al.}}",
  journal = "JINST",
  volume = "10, 9",
  pages = "C10036",
  year = "2014"
  
}

@article{EIC,
  author = "A.~Accardi, {\emph{et al.}}",
  journal = "arXiv:1212.1701 [nucl-ex]",
  note = "(2014)."
}

@manual{Apical,
 organization = "Kaneka Texas Corporation", 
 title = "\emph{Comparison of APICAL and a Competitor}", 
 url = "<http://www.kanekatexas.com/Apical\_comparison.html>."
}

@manual{Apical_Resistivity,
 organization = "Kaneka Texas Corporation", 
 title = "\emph{Typical Properties}", 
 url ="<http://www.elecdiv.kaneka.co.jp/english/apical/poli_spec.html>."
}

@article{Kapton,
  organization = "Dupont", 
  title = "\emph{Kapton Summary of Properties}",
  url = "<http://www.dupont.com/content/dam/assets/products-and-services/membranes-films/assets/DEC-Kapton-summary-of-properties.pdf>."
}


\section*{References}

\begin{thebibliography}{1}

\bibitem{Sauli:1997qp}
F.~Sauli, Nucl.\ Instrum.\ Meth.\ A 386, (1997) 531. 

\bibitem{Altunbas:2002ds}
M.C.~Altunbas and {\emph{et al.}}, Nucl.\ Instrum.\ Meth.\ A 490, (2002) 177.

\bibitem{Surrow:2010zza}
B.~Surrow, Nucl.\ Instrum.\ Meth.\ A 617, (2010) 196.

\bibitem{Gasik:2014sga}
P.~Gasik, JINST 9, (2014) C04035.

\bibitem{SBS}
K.~Gnanvo and N.~Liyanage and V.~Nelyubin and K.~Saenboonruang and Seth~Sacher and B.~Wojtsekhowski, arXiv:1409.5393 (2014).

\bibitem{Abbaneo:2014}
D.~Abbaneo and {\emph{et al.}}, JINST 10, 9 (2014), C10036.

\bibitem{EIC}
A.~Accardi and {\emph{et al.}}, arXiv:1212.1701 [nucl-ex] (2014).

\bibitem{Alfansi:2010}
M.~Alfansi and {\emph{et al.}}, Nucl.\ Instrum.\ Meth.\ A 617, (2010) 151.

\bibitem{Villa:2010w}
M.~Villa and S.D.~Pinto and M.~Alfonsi and I.~Brock and G.~Croci and E.~David and R. de Oliveria and L.~Ropelewski {\it{et al.}}, Nucl.\ Instrum.\ Meth.\ A 628, (2011) 182.

\bibitem{Tech-Etch}
Tech-Etch Inc., 45 Aldrin road, Plymouth, MA 02360, <http://www.tech-etch.com/>.

\bibitem{Becker:2006}
U.~Becker and B.~Tamm and S.~Hertel, Nucl.\ Instrum.\ Meth.\ A 556, (2006) 527.

\bibitem{Surrow:2007}
B.~Surrow and {\emph{et al.}}, Nucl.\ Instrum.\ Meth.\ A 572, (2007) 201.

\bibitem{Simon:2009}
F.~Simon and {\emph{et al.}}, Nucl.\ Instrum.\ Meth.\ A 598, (2009) 432.

\bibitem{Apical_Resistivity}
Kaneka Texas Corporation, {\emph{Typical Properties}}, http://www.elecdiv.kaneka.co.jp/english/apical/poli\_spec.html.

\bibitem{Matlab}
The Math Works Inc., Natick, MA, http://www.mathworks.com/.

\bibitem{Kapton}
Dupont, {\emph{Kapton Summary of Properties}}, http://www.dupont.com/content/dam/assets/products-and-services/membranes-films/assets/DEC-Kapton-summary-of-properties.pdf.

\bibitem{Apical}
Kaneka Texas Corporation, {\emph{Comparison of APICAL and a Competitor}}, http://www.kanekatexas.com/Apical\_comparison.html.

\bibitem{Simon:2007sk}
Simon, F. and Azmoun, B. and Becker, U. and Burns, L. and Crary, D. and {\emph{et al}}, IEEE Trans.Nucl.Sci. 54, (2007) 2646.

\end{thebibliography}
\end{document}